\documentclass[pra, aps, twocolumn, superscriptaddress]{revtex4-2}

\usepackage{slashed}
\usepackage{graphicx}
\usepackage{comment}
\usepackage[usenames, dvipsnames]{color}
\usepackage{graphics}
\usepackage{hyperref}
\usepackage{appendix}
\usepackage{bm}
\usepackage{amsfonts}
\usepackage{amsmath}
\usepackage{soul}

\hypersetup{
colorlinks=true,
urlcolor=blue,
linkcolor=blue,
linktoc=page,
citecolor=blue}

\begin{document}

\title{Plasmon-Enabled High-Precision Single Molecule Localization Microscopy over an Extended Field of View}

\author{Muzzamal I. Shaukat}
\thanks{These authors contributed equally to this work.}
\affiliation{Institute for Quantum Science and Engineering}
% \email{..........}
\author{Carlos E. Rodriguez}
\thanks{These authors contributed equally to this work.}
\affiliation{Department of Physics and Astronomy\\
Texas A\&M University, College Station, Texas 77843, USA}
\author{Junfeng Zhou}
\affiliation{Department of Physics and Astronomy\\
Texas A\&M University, College Station, Texas 77843, USA}
% \email{..........}
\author{M. Suhail Zubairy}
\email{zubairy@tamu.edu}
\affiliation{Institute for Quantum Science and Engineering}
\affiliation{Department of Physics and Astronomy\\
Texas A\&M University, College Station, Texas 77843, USA}
\author{Oumeng Zhang}
\email{ozhang@tamu.edu}
\affiliation{Institute for Quantum Science and Engineering}
\affiliation{Department of Physics and Astronomy\\
Texas A\&M University, College Station, Texas 77843, USA}

\begin{abstract}
We propose PIFLUX, a single-molecule localization scheme combining deep-subwavelength plasmonic illumination with widefield detection. A directional gap plasmon interferes with a tilted, polarization-matched reference beam, forming phase-tunable fringes, as confirmed by full-wave simulations. A Cram\'er-Rao analysis shows that, with couplers on opposite edges of the field of view, PIFLUX reaches few-nanometer precision over several micrometers, nearly doubling the precision of SIMFLUX, surpassing MINFLUX with a fixed scan pattern, and approaching iterative MINFLUX.
A maximum-likelihood estimator confirms the advantage over SIMFLUX on simulated nuclear pore complexes and on fibrils with experimental blinking statistics.
\end{abstract}

\maketitle

Single-molecule localization microscopy (SMLM) \cite{Storey1992, Storey1993, Kunze1994, Herkommer1997, Hess2006,Rust2006,Betzig2006,Sharonov2006,lelek_single-molecule_2021} has become an indispensable tool in biophysics and biochemistry, enabling access to cellular architecture, protein interactions, and molecular dynamics at spatial scales beyond the classical diffraction limit. By stochastically activating sparse subsets of fluorophores and localizing their emission, SMLM overcomes diffraction and enables molecular-scale mapping. In conventional SMLM, the localization precision scales as $\lambda/(\text{NA}\sqrt{N})$ \cite{rieger_lateral_2014}, where $\lambda$ is the emission wavelength, NA the numerical aperture, and $N$ the number of detected photons. Pushing the precision into the few-nanometer regime thus demands large photon budgets per localization, which in practice are limited by photobleaching and fluorophore blinking kinetics.

Patterned-illumination approaches have recently shown that localization precision can be improved beyond this conventional scaling. A prominent example is MINFLUX \cite{Balzarotti2017,Gwosch2020,sahl_direct_2024}, which excites single molecules (SMs) with a donut-shaped excitation beam. 
By inferring the molecular position from photon counts acquired at multiple beam positions, MINFLUX can reach $\sim1$~nm precision with substantially fewer photons than centroid-based localization. 
However, MINFLUX probes a region of only $\sim$50--100~nm around one emitter at a time, so a larger field of view (FOV) is imaged by addressing emitters sequentially, which limits throughput.

To mitigate this limitation, SIMFLUX combines sinusoidal illumination with widefield camera detection \cite{Cnossen2020}. By correlating emitter positions with photon counts across phase-shifted pattern exposures, SIMFLUX enables patterned-illumination localization over fields of view of tens of micrometers. Yet its precision gain is fundamentally limited by the illumination-pattern pitch; for standing-wave interference this is bounded by $\sim\lambda/(2\text{NA})$, restricting the improvement to roughly a twofold enhancement relative to conventional SMLM.

This motivates a central question: can the illumination-pattern pitch be reduced further to approach MINFLUX-level localization precision while retaining the large field of view and parallelism of widefield detection?
Achieving a smaller pitch requires optical fields with larger transverse spatial frequencies than are available in conventional far-field interference. 
Beyond fluorescence microscopy, structured optical fields have long been exploited for subwavelength localization across other areas of physics. In particular, Zubairy and collaborators developed a broad range of quantum-optical approaches to subwavelength localization and imaging based on coherent atom-field interactions and quantum interference, demonstrating spatial resolution beyond the classical diffraction limit \cite{Qamar2000,Azim2004,Sahrai2005,Kapale2006,Macovei2007,Kiffner2008, Chang2006, Hebin2008,Qamar2009, Liao2012,Sun2011}.
Drawing inspiration from these approaches, we exploit the large in-plane wavevectors of evanescent plasmonic near fields to generate deep-subwavelength illumination patterns over an extended field of view.

Surface plasmon polaritons (SPPs) are evanescent electromagnetic modes bound to metal-dielectric interfaces, central to plasmonic control of light at the nanoscale
\cite{maier_plasmonics_2007,economou_surface_1969,raether_surface_1988,berini_surface_2012}. Their guided propagation and strong near-field confinement support optical fields with in-plane wavevectors exceeding those of freely propagating light, thereby providing access to spatial frequencies beyond the diffraction limit
\cite{barnes_surface_2003,bozhevolnyi_channel_2006,gramotnev_plasmonics_2010}. This capability enables optical dispersion and field localization to be engineered through material composition, operating frequency, and geometry
\cite{maier_plasmonics_2007,barnes_surface_2003,schuller_plasmonics_2010}. In multilayer plasmonic systems, coupling between neighboring metal-dielectric interfaces gives rise to hybridized modes with tunable confinement, propagation length, and field enhancement, motivating extensive studies of multilayer architectures for nanoscale guiding, imaging, sensing, and light--matter interactions
\cite{dionne_highly_2006,wickremasinghe_controlling_2015,liu_one-dimensional_2020,saeidi_design_2022,verma_review_2024}.

The present water-metal-dielectric-metal-water (WMDMW) structure belongs to the broader class of finite-thickness metal-dielectric-metal (MDM) plasmonic waveguides, which support strongly confined gap-plasmon modes within nanoscale dielectric spacers. The coupling between the two metal-dielectric interfaces gives rise to even- and odd-symmetry gap-plasmon modes, whose dispersion and confinement can be tuned through the thickness of the spacer, the thickness of the metal, and the surrounding dielectric environment \cite{economou_surface_1969,burke_surface-polariton-like_1986,yang_long-range_1991,verhagen_near-field_2008,xiang_long-range_2013}. As the dielectric gap decreases, the modal confinement increases and the in-plane propagation constant \(\beta'\) becomes substantially larger than the free-space wavevector, enabling deeply subwavelength optical fields and illumination patterns
\cite{maier_plasmonics_2007,bozhevolnyi_channel_2006,gramotnev_plasmonics_2010}. These characteristics have established MDM architectures as a versatile platform for nanophotonic integration, subwavelength waveguiding, and plasmonic circuitry \cite{dionne_plasmon_2006,messner_plasmonic_2023,zhang_hybrid_2017,zhang_chip_2024,rojas_yanez_plasmonic_2025}. In realistic finite-thickness structures, both the surrounding dielectric medium and the metal-film thickness influence the modal dispersion and propagation loss of the supported gap-plasmon modes
\cite{economou_surface_1969,burke_surface-polariton-like_1986,yang_long-range_1991}. 
Plasmonic structured illumination has previously been proposed and
experimentally demonstrated for super-resolution fluorescence imaging
using both propagating surface plasmons and localized plasmonic fields
\cite{Wei2010,Wei2014,Ponsetto2014,Ponsetto2017}. Experimental PSIM and LPSIM implementations reconstruct
super-resolution images from multiple fluorescence
images acquired under different plasmonic illumination
patterns \cite{Wei2014,Ponsetto2017}. Here, we incorporate plasmonic
illumination into a single-molecule localization framework, in which
photon counts recorded under successive excitation phases provide
information about individual emitter positions. Our contribution is the
formulation and analysis of this plasmon-assisted localization scheme.

\begin{figure}[t!]
  \centering
  \includegraphics[scale=1]{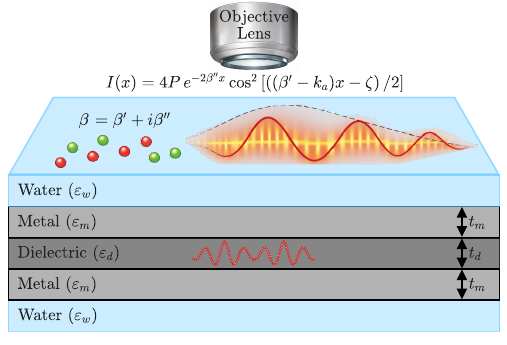}
  \caption{Surface plasmon polariton mode supported by a finite water--metal--dielectric--metal--water structure. The complex propagation constant $\beta = \beta' + i\beta''$ determines the spatial oscillation and attenuation of the gap‑plasmon field, which excites nearby fluorescent emitters whose emission is collected by an objective lens.}
  \label{fig:ND}
\end{figure}

We term this method PIFLUX (plasmon-illumination FLUX), in analogy with other structured-illumination localization schemes. 
We first introduce a WMDMW architecture that supports SPP excitation patterns with pitches substantially smaller than those attainable with conventional optics. Using a Cram\'er-Rao bound analysis, we show that PIFLUX can approach MINFLUX-level localization precision over a field of view spanning several micrometers. 
Finally, we develop a position estimator for SM localizations and validate it on simulated nuclear pore complexes and on fibrils with experimental blinking statistics.

We consider a finite-thickness water--metal--dielectric-- metal--water structure, where a dielectric spacer of thickness $t_d$ is confined between two metal films of thickness $t_m$ (Fig. \ref{fig:ND}). The supported gap-plasmon modes are obtained from the transverse-magnetic dispersion relation of the multilayer geometry, which explicitly incorporates the finite metal thickness and surrounding dielectric environment. Numerical solution yields a complex propagation constant $\beta=\beta'+i\beta''$, where $\beta'$ determines the plasmonic spatial frequency and $\beta''$ characterizes propagation loss.

\begin{figure}[t!]
    \centering
    \includegraphics{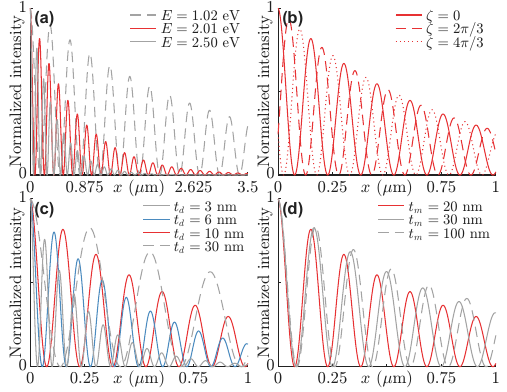}
    \caption{Normalized intensity produced by interference between a directional gap plasmon and a coherent applied field, computed from Eq.~\eqref{eqn:Final_Intensity} under amplitude, attenuation, and polarization matching. Each pattern is normalized by the reference intensity $4P(z_0)$, with $\phi=\chi+\zeta$. (a) Dependence of the gap-plasmon wavevector on excitation energy. Increasing energy increases $\beta'$ and reduces the modulation period. (b) Phase-dependent shift of the interference pattern. (c) Dependence of the plasmonic intensity modulation on spacer thickness. Smaller gaps increase $\beta'$ and enhance confinement.
    (d) Dependence on metal thickness. Thinner metal films increase both $\beta'$ and the attenuation.}
    \label{fig:Energy}
\end{figure}
A directional gap plasmon launched at $x=0$ propagates along
$+x$ with longitudinal dependence $e^{i\beta'x}e^{-\beta''x}$.
Its field extends through the finite upper metal film into water, where it excites the fluorophores at a height $a$ above the upper metal surface ($z_0=v+a$, Supplemental Sec.~A).
Now consider simultaneous illumination by a coherent applied
field
$\mathbf{E}_a
=
E_0g_a(x)e^{i(k_a x+\phi)}
\left(
\cos\theta\,\hat{\mathbf{x}}
+
e^{i\gamma}\sin\theta\,\hat{\mathbf{z}}
\right)$,
where $g_a(x)$ is the amplitude envelope, $k_a$ determines
the phase variation along $x$, and $\phi$ is the controlled
relative phase. The parameters $\theta$ and $\gamma$ specify
the relative amplitudes and phase of the polarization components.
The applied field is provided by a $p$-polarized beam incident from the water side at angle
$\alpha$, so that $k_a=k_0\sqrt{\varepsilon_w}\sin\alpha$. The parameters $\theta$ and $\gamma$ are
then fixed by $\alpha$ through the reflection of this beam from the multilayer.
The total excitation intensity, derived in Supplemental Sec.~B, is 
\begin{eqnarray}
I_{\mathrm{tot}}(x,z_0)
&=&
E_0^2g_a^2(x)+P(z_0)e^{-2\beta''x}+2E_0Q(z_0)g_a(x)
\nonumber\\
&&
e^{-\beta''x}
\cos\left[
(\beta'-k_a)x-\phi+\chi
\right].
\label{eqn:Final_Intensity}
\end{eqnarray}
Here, $P(z_0)$ is the squared local SPP amplitude, while
$Q(z_0)$ and $\chi(z_0)$ are the magnitude and phase of
its polarization overlap with the applied field, as defined
in Supplemental Sec.~B. We consider the ideal matched case
$g_a(x)=e^{-\beta''x}$,
$E_0=\sqrt{P(z_0)}$, and $Q(z_0)=\sqrt{P(z_0)}$,
which permits exact intensity nodes.
The ideal model assumes exact polarization matching, $Q=\sqrt{P}$; in the device, the applied beam is tilted along the SPP propagation direction at the angle that maximizes the overlap, which satisfies this condition approximately (Supplemental Sec.~B).
Interference between the incident beam and its reflection from the multilayer then shifts $E_z$ by nearly $\pi/2$ relative to $E_x$, so that the applied field approximately matches the elliptical polarization of the SPP field, $E_z/E_x=i\beta/\kappa_w$.
Spatial modulation is
governed by the wavevector difference $\beta'-k_a$, giving
a fringe period $\Lambda=2\pi/|\beta'-k_a|$.
Because matching requires $k_a>0$, $\Lambda$ is slightly larger than $2\pi/\beta'$.
Varying $\phi$ shifts the interference fringes while leaving
the attenuation envelope unchanged.
Dielectric spacers of 6 and 10~nm are selected as two tradeoffs between strong confinement
and long propagation length, consistent with the meta-sandwich geometry of Ref.~\cite{cao_numerical_2017}.
The metal permittivity is taken from the experimentally tabulated optical constants of Ref.~\cite{johnson_optical_1972}.

Fig.~\ref{fig:Energy}a shows that increasing the excitation energy increases $\beta'$ and compresses the fringe period $\Lambda$, reflecting the shortening of the plasmon wavelength and establishing energy-tunable subwavelength structured illumination. Fig.~\ref{fig:Energy}b shows that varying $\zeta$ shifts the interference pattern while preserving $\Lambda$; this phase-controlled translation enables structured illumination without altering the underlying plasmon wavelength. As shown in Fig.~\ref{fig:Energy}c, decreasing the spacer thickness strengthens the coupling between the two metal--dielectric interfaces, increasing $\beta'$ and reducing $\Lambda$, at the cost of stronger attenuation. Fig.~\ref{fig:Energy}d shows that thinner metal films likewise increase both $\beta'$ and the attenuation $\beta''$, so that the fringes decay more rapidly.

\begin{figure}[t!]
    \centering
    \includegraphics{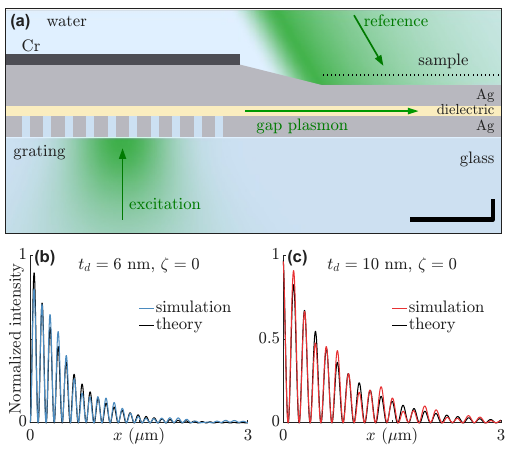}
    \caption{Full-wave simulation of PIFLUX. (a) Cross section of the simulated structure for $t_d = 10$~nm. A line-focused Gaussian excitation beam (500~nm FWHM) incident from the glass substrate is coupled by a ten-period grating of glass-filled slits in the lower Ag film into a gap plasmon. A Cr-capped Ag overlayer, tapered over 500 nm, covers the grating region and blocks the excitation light from reaching the sample. A $p$-polarized reference beam is incident from the water side at the polarization-matched angle ($\alpha = 29.2^\circ$ and $30.1^\circ$ for $t_d = 6$ and 10 nm), with amplitude envelope $e^{-\beta'' x}$. The reference illuminates the sample region starting at the end of the taper ($x = 0$), i.e., assuming a field stop. The dotted line marks the sample plane, $a = 10$ nm above the metal surface. Scale bars: 500 nm horizontal, 20 nm vertical. (b,c) Normalized excitation intensity at the sample plane for $\zeta = 0$ from finite-element simulations (color) and Eq.~\eqref{eqn:Final_Intensity} (black), for (b) $t_d = 6$ nm (grating period 99 nm, slit width 35 nm) and (c) $t_d = 10$ nm (period 130 nm, slit width 50 nm).}
    \label{fig:full_wave}
\end{figure}

To demonstrate that a physical system can generate this pattern, we performed two-dimensional full-wave finite-element simulations of a realistic device, assumed invariant along $y$ (Fig.~\ref{fig:full_wave}a).
For experimental realism, the water below the structure is replaced by a glass substrate, which shifts $\beta'$ by only 0.3\% and 0.5\% for $t_d = 6$ and 10~nm, respectively. A line-focused Gaussian excitation beam incident from the substrate is coupled into a gap plasmon by a grating of glass-filled slits in the lower Ag film, with a period equal to the gap-plasmon wavelength $2\pi/\beta'$. A tapered, Cr-capped Ag overlayer covers the grating and prevents the transmitted excitation light from reaching the sample. The reference beam is incident from the water side at the polarization-matched angle and is confined to the sample region by a field stop.

The simulated excitation patterns closely follow Eq.~\eqref{eqn:Final_Intensity} (Figs.~\ref{fig:full_wave}b,c and S1). For $t_d = 6$~nm, the fringe visibility averages 0.96 and the fringe positions deviate from the model by 4~nm (rms) for $x \leq 2~\mu$m. Beyond this distance, the gap plasmon has decayed to below 2\% of its initial intensity, and residual surface waves launched at the coupler reduce the visibility. For $t_d = 10$~nm, the longer propagation length maintains an average visibility of 0.96, with a 7~nm rms fringe-position deviation, over the full 3~$\mu$m range. These deviations could be reduced further, for example by optimizing the thickness and profile of the Ag cap, which would extend the range over which the ideal pattern is maintained. 
Cross-sectional field maps and a three-dimensional simulation of the corner where the $x$- and $y$-oriented couplers meet are provided in Supplemental Sec.~C (Figs.~S1 and S2).

\begin{figure}[t!]
    \centering
    \includegraphics[width=\linewidth]{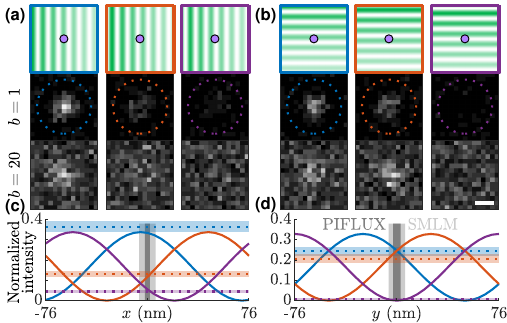}
    \caption{Principle of PIFLUX. (a,b) Phase-shifted excitation patterns (green) for $t_d=10$~nm and 2.01~eV ($\lambda_{\mathrm{ex}}=617$~nm), translated along (a) $x$ and (b) $y$, together with representative synthetic camera frames. The purple dot marks the SM position. Six images are acquired per localization (one-sided excitation). The total signal is 500 photons and the total background is 1 (middle row) or 20 (bottom row) photons per $(65~\mathrm{nm})^2$ pixel (summed over all six frames). Scale bar: 282~nm ($\lambda_{\mathrm{em}}/2\mathrm{NA}$). (c,d) Photon counts for the three phase shifts compared with the corresponding excitation profiles along (c) $x$ and (d) $y$. Solid curves: normalized excitation intensity; dotted curves: normalized detected emission, with shaded areas indicating intensity uncertainty; light and dark gray areas: estimated position with its uncertainty for SMLM and PIFLUX, respectively.}    \label{fig:principle}
\end{figure}

We next describe how the SPP intensity pattern enables PIFLUX; the concept is illustrated in Fig.~\ref{fig:principle}. 
For each localization event, six images are acquired: three with the excitation pattern phase shifted along $x$ (Fig.~\ref{fig:principle}a) and three along $y$ (Fig.~\ref{fig:principle}b), with successive phase shifts of $2\pi/3$ in each direction. 
Using two orthogonal directions is sufficient to obtain nearly isotropic localization precision \cite{Cnossen2020}, in contrast to Fourier-space filling approaches that require multiple pattern orientations (e.g., conventional \cite{gustafsson_surpassing_2000}, nonlinear~\cite{gustafsson_nonlinear_2005} and plasmon-enabled \cite{abbas_deep_2018} structured illumination microscopy). 
In each direction, successive exposures are phase shifted by $\Delta\zeta=2\pi/3$, so the detected photon counts vary across the three frames according to the molecule's position within the pattern. 
By first assigning the emitter to a single period of the decaying sinusoidal profile and then refining its position by matching the measured counts to the known excitation patterns, PIFLUX extracts additional positional information and achieves improved localization precision relative to conventional SMLM (Fig.~\ref{fig:principle}c,d).

\begin{figure}[t!]
    \centering
    \includegraphics[width=\linewidth]{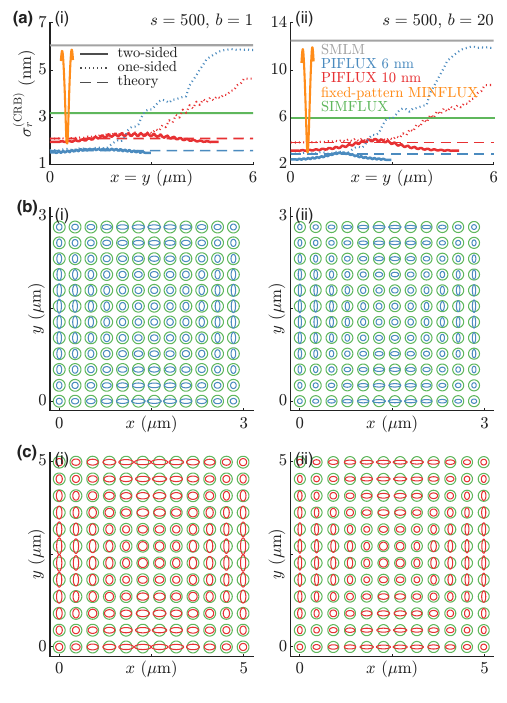}
    \caption{Lateral localization CRB for 500 signal photons and background levels of (i) 1 and (ii) 20 photons per (65~nm)$^2$ pixel (summed over all frames). (a) Mean localization CRB $\sigma_r^{(\mathrm{CRB})}$ along the diagonal $x=y$ for two-sided (solid) and one-sided (dotted) excitation, computed from the full-wave excitation patterns, together with the theoretical model [Eq.~\eqref{eqn:Final_Intensity}] for one-sided excitation (dashed). All curves are averaged over the fringe phase with a sliding window of three fringe periods (one period for SIMFLUX). For two-sided excitation, the curves end at the coupler spacing $D=3$~\textmu{}m ($t_d=6$~nm) and 5~\textmu{}m ($t_d=10$~nm). Gray: conventional SMLM; blue: PIFLUX with $t_d=6$~nm; red: PIFLUX with $t_d=10$~nm; orange: fixed-pattern MINFLUX with a donut shift diameter of 100~nm centered at $x=0.5$~\textmu{}m; green: SIMFLUX. (b, c) CRB covariance ellipses for two-sided excitation over the $D\times D$ field of view for (b) $t_d=6$~nm and (c) $t_d=10$~nm, from the full-wave patterns, compared with SIMFLUX. Each ellipse is averaged over one-third of a fringe period around the grid point. The ellipses are magnified by factors of 30 and 15 in (b) and 50 and 25 in (c) for (i) and (ii), respectively.}
    \label{fig:CRB}
\end{figure}

To benchmark PIFLUX against other techniques, we computed the Cram\'er-Rao bound (CRB) from the Fisher information (FI) \cite{moon2000mathematical,chao_fisher_2016} of the Poisson-distributed photon counts, using a scalar point-spread function (PSF, see Supplemental Sec. D). We fixed the excitation energy at 2.01 eV ($\lambda_\text{ex}=617$~nm), metal thickness $t_m=20$~nm, and emission wavelength $\lambda_\text{em}=676$~nm, with a water-immersion objective of $\text{NA} = 1.2$ ($n=1.33$) for all methods, since the emitters are imaged from the water side. For PIFLUX, we used the full-wave excitation patterns of Fig.~\ref{fig:full_wave} and the ideal model of Eq.~\eqref{eqn:Final_Intensity}. In the one-sided scheme described above, in which the gap plasmon is launched from a single coupler, the gap plasmon decays with distance from the coupler, so that residual surface waves eventually dominate the excitation and reduce the fringe visibility (Fig.~\ref{fig:full_wave}). We therefore also consider a two-sided scheme, in which identical couplers at both edges of each axis ($x=0,D$ and $y=0,D$) are launched in turn, with the reference beam tilted along the respective propagation direction, so that every position lies within $D/2$ of a coupler; we use $D=3$~\textmu{}m for $t_d=6$~nm and 5~\textmu{}m for $t_d=10$~nm.
Reflection of the plasmon at the opposite coupler has a negligible effect: even for complete reflection, the CRB changes by only $\sim0.1$\%, because the reflected plasmon has nearly orthogonal polarization and does not interfere with the reference beam, and it is appreciable only near the opposite coupler, where most photons are recorded in the frames launched from that coupler.
The total signal $s$ and background $b$ per pixel are distributed over $N$ frames ($N=6$ for one-sided PIFLUX and SIMFLUX, 12 for two-sided PIFLUX, 4 for MINFLUX, 1 for SMLM), so all methods are compared for the same number of detected photons; for PIFLUX, emitters far from the couplers, where the excitation is weakest, need correspondingly longer to emit them (Supplemental Sec.~D).
For MINFLUX, we assume a residual intensity of 0.5\% of the peak at the donut zero.

The resulting CRBs are shown in Fig.~\ref{fig:CRB}. We define an overall lateral localization precision, $\sigma_r^{(\mathrm{CRB})}=\det(\mathrm{CRB})^{1/4}$. 
This metric is the geometric mean of the standard deviations along the principal axes of the error ellipse, which reduces to that of the precisions along $x$ and $y$ when the $x$--$y$ covariance is zero.
We evaluated the CRBs along the diagonal $x=y$, where the frames of the two axes receive equal shares of the photons and $\sigma_x^{(\mathrm{CRB})}=\sigma_y^{(\mathrm{CRB})}$ by symmetry (Fig.~\ref{fig:CRB}a), and over the two-dimensional FOV (Fig.~\ref{fig:CRB}b,c). Compared with conventional SMLM, PIFLUX provides better localization precision throughout the FOV under both low and high signal-to-background ratio (SBR) conditions.

Our main comparison, however, is between PIFLUX and SIMFLUX. Because of the excitation symmetry in PIFLUX and SIMFLUX, the localization precision varies within each one-third period of the decaying sinusoidal excitation pattern by about $\pm 20\%$. We therefore show the CRB averaged over the fringe phase within a sliding window. With one-sided excitation, PIFLUX with $t_d=6$~nm reaches 1.6~nm ($b=1$) and 3.0~nm ($b=20$) near the coupler, in agreement with the ideal model, and remains better than SIMFLUX up to $x\approx 3$~\textmu{}m (Fig.~\ref{fig:CRB}a). Beyond this distance, the residual surface waves reduce the fringe visibility and the precision approaches that of SMLM.
Two-sided excitation removes this limitation: along the diagonal, $\sigma_r^{(\mathrm{CRB})}$ stays within 1.5--1.7~nm ($b=1$) and 2.4--3.0~nm ($b=20$) over the full 3~\textmu{}m spacing, nearly twice the precision of SIMFLUX (3.2 and 5.9~nm). 
It averages 1.8 and 3.0~nm over the $3\times3$~\textmu{}m$^2$ FOV and is better than that of SIMFLUX at every sampled position (Fig.~\ref{fig:CRB}b and Supplemental Sec.~D).
Notably, two-sided PIFLUX with $t_d=6$~nm is even 21\% better than MINFLUX with a donut shift diameter of $100$~nm at high SBR [Fig.~\ref{fig:CRB}a(i)] and 14\% at low SBR [Fig.~\ref{fig:CRB}a(ii)]. Iterative MINFLUX, which re-centers the donut and shrinks the scan diameter (100, 50, and 25~nm, with one third of the photons each), reaches 1.0 and 1.6~nm under the same SBRs, but it localizes one molecule at a time and requires real-time feedback.

For PIFLUX with $t_d=10$~nm, which may be easier to manufacture experimentally, the overall trend remains the same, although the improvement over SIMFLUX is smaller.
In this case, the improvement in overall localization precision along the diagonal is 50\% ($b=1$) and 69\% ($b=20$) for two-sided excitation. However, its longer propagation length extends the one-sided range in which PIFLUX outperforms SIMFLUX to $x\approx4$~\textmu{}m (Fig.~\ref{fig:CRB}a), and allows a larger coupler spacing of 5~\textmu{}m for two-sided excitation, over which $\sigma_r^{(\mathrm{CRB})}$ (2.2 and 3.6~nm on average) remains better than that of SIMFLUX at every sampled position (Fig.~\ref{fig:CRB}c).
These results show that PIFLUX offers a favorable tradeoff between localization precision, FOV size, and practical implementation.

\begin{figure}[b!]
    \centering
    \includegraphics[width=\linewidth]{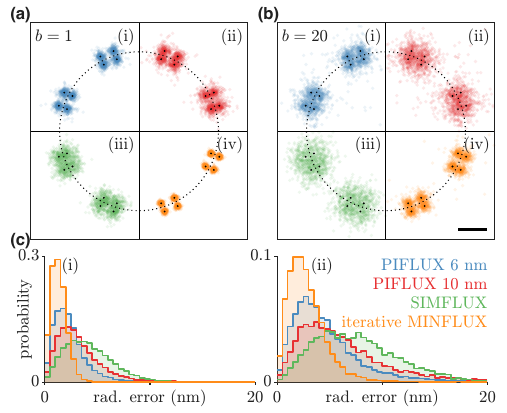}
    \caption{Localization of simulated nuclear pore complexes (NPCs). Localizations of 250 Nup96 NPCs (32 copies on a 110-nm-diameter ring) at random positions and orientations across the FOV, aligned to a common NPC frame, for 500 photons per localization and a total background of (a) $b=1$ and (b) $b=20$ photons per pixel. Each quadrant shows one method: (i) PIFLUX with $t_d=6$~nm, (ii) PIFLUX with $t_d=10$~nm, (iii) SIMFLUX, and (iv) iterative MINFLUX. Dotted circle: NPC ring; black dots: true copy positions. Scale bar: 20~nm. (c) Distribution of the radial localization error for (i) $b=1$ and (ii) $b=20$.}
    \label{fig:npc}
\end{figure}

Another advantage of the two-sided scheme is that it can be tiled across a chip, because a symmetric grating launches gap plasmons in both directions. Couplers placed on a square grid with period $D$ are excited in two alternating sets, so that in each set every $D\times D$ cell is illuminated from one of its two edges, and the two sets together provide two-sided excitation for all cells simultaneously, with the same twelve frames per localization. Adjacent cells then require reference beams tilted in opposite directions, which can be supplied by two reference beams confined to alternate cells with a patterned field stop, for example a digital micromirror device in a conjugate image plane.
The plasmon leaking into the neighboring cell is attenuated by at least $e^{-2\beta''D}$ ($\sim$0.2\% for $t_d=6$~nm with $D=3$~\textmu{}m). Therefore, apart from the strips occupied by the couplers, the field of view is limited by the patterned area rather than by the plasmon propagation length.

%%%%%% Estimator section  %%%%%%

Finally, we evaluated PIFLUX with a maximum-likelihood estimator on simulated data and benchmarked it against SIMFLUX and MINFLUX under the conditions of the CRB analysis (Supplemental Sec.~E). The estimator first locates each emitter using the sum of its frames and then maximizes the Poisson likelihood of the photon counts collected in a $7\times7$-pixel region of interest (ROI) for each axis. As a test structure, we used Nup96 nuclear pore complexes (NPCs), projected in 2D as 32 copies at the eight corners of a 110-nm-diameter ring, with 250 NPCs at random positions and orientations across the FOV. As in the CRB analysis, each localization used 500 photons and a total background of $b=1$ or 20 photons per pixel, and SIMFLUX and iterative MINFLUX ($L=100$, 50, and 25~nm, with the first iteration centered on the emitter) localized the same NPCs.

Figure~\ref{fig:npc} shows the localizations of all NPCs, aligned to a common NPC frame. 
At $b=1$, PIFLUX resolves the substructure of each corner for both spacer thicknesses, which SIMFLUX does not.
The median radial errors are 2.1~nm for PIFLUX with $t_d=6$~nm, 2.9~nm with $t_d=10$~nm, 3.9~nm for SIMFLUX, and 1.3~nm for MINFLUX. At $b=20$, they increase to 3.9, 5.5, 7.1, and 2.6~nm, respectively.

We further tested PIFLUX on $\alpha$-synuclein fibrils, taking the emitter positions and photon counts from an experimental PAINT dataset~\cite{bruggeman_polcam_2024} and generating the blink on-times from emission rates sampled from the same data (Supplemental Sec.~E). At the same number of detected photons, the median localization error of PIFLUX with $t_d=6$~nm is lower than that of SIMFLUX above $\sim$300 photons, by up to a factor of 1.8.

%%%%%% Summary section  %%%%%%

In summary, we have developed an analytical model for the interference between a gap plasmon in a finite metal--dielectric--metal structure and a tilted, polarization-matched reference beam, and used it to introduce PIFLUX, a localization scheme that combines deep-subwavelength plasmonic illumination with widefield detection. Full-wave simulations of a grating-coupled device reproduce the predicted fringes.
For the same number of detected photons, a Cram\'er-Rao analysis shows that,
with couplers on opposite edges of the field of view, PIFLUX reaches few-nanometer localization precision over several micrometers, nearly twice the precision of SIMFLUX, higher than that of MINFLUX with a fixed scan pattern, and approaching that of iterative MINFLUX, while retaining the parallelism of camera-based detection; tiling the couplers across a chip extends the field of view further.
A maximum-likelihood estimator resolves the corner substructure of simulated nuclear pore complexes, and, on fibrils with experimental blinking statistics, achieves lower median errors than SIMFLUX at the same number of photons. By decoupling localization precision from the diffraction-limited illumination pitch, PIFLUX offers a route to high-throughput fluorescence imaging with single-digit-nanometer precision without sequential, emitter-by-emitter excitation.

%\begin{acknowledgments} This work is supported by \hl{xx} \end{acknowledgments}

%%%%%%%%%%%%%%%%%%%%
\bibliography{references}

\end{document}

\title{Supplemental Material for Plasmon-Enabled High-Precision Single Molecule Localization Microscopy over an Extended Field of View}

\author{Muzzamal I. Shaukat}
\thanks{These authors contributed equally to this work.}
\affiliation{Institute for Quantum Science and Engineering}
\author{Carlos E. Rodriguez}
\thanks{These authors contributed equally to this work.}
\affiliation{Department of Physics and Astronomy\\
Texas A\&M University, College Station, Texas 77843, USA}
\author{Junfeng Zhou}
\affiliation{Department of Physics and Astronomy\\
Texas A\&M University, College Station, Texas 77843, USA}
\author{M. Suhail Zubairy}
\email{zubairy@tamu.edu}
\affiliation{Institute for Quantum Science and Engineering}
\affiliation{Department of Physics and Astronomy\\
Texas A\&M University, College Station, Texas 77843, USA}
\author{Oumeng Zhang}
\email{ozhang@tamu.edu}
\affiliation{Institute for Quantum Science and Engineering}
\affiliation{Department of Physics and Astronomy\\
Texas A\&M University, College Station, Texas 77843, USA}

\maketitle

\appendix

\counterwithout{equation}{section}
\renewcommand{\theequation}{S\arabic{equation}}

\section{Derivation of the dispersion relation for the water-clad MDM structure}
\label{app:water_mdm}

We consider a symmetric water--metal--dielectric-- metal--water structure consisting of a dielectric spacer of relative permittivity $\varepsilon_d$ and thickness $t_d$, bounded by two identical metal films of thickness $t_m$ and frequency-dependent relative permittivity $\varepsilon_m(\omega)$. The exterior dielectric regions have relative permittivity $\varepsilon_w$. We choose the origin of the transverse coordinate at the center of the dielectric spacer and define $u=t_d/2$ and $v=t_d/2+t_m$. The dielectric--metal interfaces are therefore located at $z=\pm u$, and the outer metal--water interfaces are located at $z=\pm v$.
The relative-permittivity distribution is
\begin{equation}
\varepsilon(z)=
\begin{cases}
\varepsilon_d, & |z|<u,\\
\varepsilon_m(\omega), & u<|z|<v,\\
\varepsilon_w, & |z|>v.
\end{cases}
\label{eq:A_permittivity}
\end{equation}
Here, $\varepsilon(z)$ denotes the position-dependent relative permittivity of the symmetric five-layer structure. The metal films have finite thickness, whereas the exterior water regions are treated as semi-infinite. We adopt the harmonic time dependence $e^{-i\omega t}$ and treat the constituent media as linear, isotropic, and nonmagnetic within a local-response model.
The corresponding constitutive relations are $\mathbf D=\varepsilon_0\varepsilon(z) \mathbf E$ and $\mathbf B=\mu_0\mathbf H$.
The $x$ axis is chosen along the SPP propagation direction and the $z$ axis is normal to the planar interfaces. 
For TM polarization, the nonvanishing field components are $H(x,z)=\hat{\mathbf y}H_y(x,z)$ and $E(x,z)=\hat{\mathbf x}E_x(x,z)+\hat{\mathbf z}E_z(x,z)$.
For $\mathbf H=(0,H_y,0)$, Maxwell's equations ($\nabla\times\mathbf E=i\omega\mu_0\mathbf H$, $\nabla\times\mathbf H
=-i\omega\varepsilon_0\varepsilon(z)\mathbf E$) give
\begin{align}
E_x(x,z)
&=-\frac{i}{\omega\varepsilon_0\varepsilon(z)}
\frac{\partial H_y(x,z)}{\partial z},
\label{eq:A_Ex}\\
E_z(x,z)
&=\frac{i}{\omega\varepsilon_0\varepsilon(z)}
\frac{\partial H_y(x,z)}{\partial x}.
\label{eq:A_Ez}
\end{align}
Within each homogeneous region $j\in\{d,m,w\}$,
elimination of the electric field gives
\begin{equation}
\left(
\frac{\partial^2}{\partial x^2}
+\frac{\partial^2}{\partial z^2}
+\varepsilon_jk_0^2
\right)H_y(x,z)=0,
\qquad k_0=\frac{\omega}{c}.
\label{eq:A_Helmholtz}
\end{equation}
Because the multilayer is translationally invariant
along $x$, the longitudinal and transverse dependences
of each eigenmode can be separated. For a mode
propagating in the positive $x$ direction, we write
$H_y^{(+)}(x,z)=A_+h_j(z)e^{i\beta x}$,
where $\beta=\beta'+i\beta''$ is the complex
propagation constant and $h_j(z)$ is the transverse
magnetic-field profile in layer $j$.
Within each homogeneous region,
\begin{equation}
\frac{d^2h_j(z)}{dz^2}-\kappa_j^2h_j(z)=0,
\qquad
\kappa_j=\sqrt{\beta^2-\varepsilon_jk_0^2}.
\label{eq:A_mode_equation}
\end{equation}
For a bound mode, the exterior square-root branch
is chosen so that $\operatorname{Re}\kappa_w>0$.
The general solution in each finite homogeneous
layer can be written as
\begin{equation}
h_j(z)=B_j e^{\kappa_jz}+D_j e^{-\kappa_jz}.
\label{eq:A_general_solution}
\end{equation}
The coefficients are determined by transverse decay
in the exterior regions and the electromagnetic
boundary conditions at the four interfaces. Since $\varepsilon(z)=\varepsilon(-z)$, the
eigenmodes may be chosen to have definite parity.
Inside the dielectric spacer, the general solution
may equivalently be written as
$h_d(z)=B_d\cosh(\kappa_dz)+D_d\sinh(\kappa_dz)$.
For the even-$H_y$ branch, $h_d(-z)=h_d(z)$ and
hence $D_d=0$. The modal profile is determined only
up to an arbitrary multiplicative constant.
We choose $h_d(0)=1$, so that $h_d(z)=\cosh(\kappa_dz)$($|z|<u$).
In the upper water region ($z>v$), the decaying solution is $h_w(z)=B_w e^{-\kappa_w(z-v)}$.
To impose the outer boundary conditions, we write
the upper metal profile ($u<z<v$) as $h_m(z)=B_m\cosh[\kappa_m(v-z)]
+D_m\sinh[\kappa_m(v-z)]$.
Continuity of the tangential magnetic field at
$z=v$ gives $B_m=B_w$. Continuity of the tangential
electric field requires
$h_m'(v)/\varepsilon_m=h_w'(v)/\varepsilon_w$,
where $h_j'=dh_j/dz$. Since
$h_m'(v)=-\kappa_mD_m$ and
$h_w'(v)=-\kappa_wB_w$, we obtain $D_m=P_w B_w/P_m$ and $P_j=\kappa_j/\varepsilon_j$.
The upper metal profile is therefore
\begin{equation}
h_m(z)=B_w\biggl\{
\cosh[\kappa_m(v-z)]
+\frac{P_w}{P_m}\sinh[\kappa_m(v-z)]
\biggr\}.
\label{eq:A_metal_profile}
\end{equation}
Continuity of $H_y$ at $z=u$ gives $\cosh(\kappa_d u) =B_w\left[ \cosh(\kappa_mt_m) +P_w\sinh(\kappa_mt_m)/P_m \right]$.
Thus,
\begin{equation}
B_w=
\frac{\cosh(\kappa_dt_d/2)}
{\cosh(\kappa_mt_m)
+(P_w/P_m)\sinh(\kappa_mt_m)}.
\label{eq:A_Bw}
\end{equation}
Continuity of $E_x$ at the same interface gives
\begin{equation}
\begin{split}
P_d\sinh(\kappa_d u)
=-B_w\bigl[
&P_m\sinh(\kappa_mt_m)\\
&+P_w\cosh(\kappa_mt_m)
\bigr].
\end{split}
\label{eq:A_E_matching}
\end{equation}
Eliminating $B_w$ yields
\begin{equation}
\frac{\kappa_d}{\varepsilon_d}
\tanh\!\left(\frac{\kappa_dt_d}{2}\right)
+P_{\mathrm{in}}^{(m)}=0,
\label{eq:A_dispersion}
\end{equation}
where
\begin{equation}
P_{\mathrm{in}}^{(m)}
=P_m
\frac{P_w+P_m\tanh(\kappa_mt_m)}
{P_m+P_w\tanh(\kappa_mt_m)}.
\label{eq:A_admittance}
\end{equation}
This relation determines the allowed complex
propagation constant of the even-$H_y$ mode.
For completeness, the odd-$H_y$ branch has a
dielectric profile proportional to $\sinh(\kappa_dz)$
and satisfies
\begin{equation}
\frac{\kappa_d}{\varepsilon_d}
\coth\!\left(\frac{\kappa_dt_d}{2}\right)
+P_{\mathrm{in}}^{(m)}=0.
\label{eq:A_odd_dispersion}
\end{equation}
The following analysis is restricted to the
even-$H_y$ branch. The finite metal thickness and the exterior
dielectric enter through
$P_{\mathrm{in}}^{(m)}$.
When $\operatorname{Re}(\kappa_mt_m)\gg1$,
$P_{\mathrm{in}}^{(m)}\rightarrow P_m$,
and Eq.~\eqref{eq:A_dispersion} reduces to the
three-layer MDM dispersion relation. 

We consider an SPP excited near $x=0$ through
coupling to the dielectric spacer.
The magnetic field
of a directional mode is
$H_{y,\sigma}(x,z)
=A_\sigma h(z)e^{i\sigma\beta x}$ ($\sigma=\pm1$), where $A_\sigma$ is the complex magnetic-field amplitude and $\sigma=\pm 1$ denotes propagation along $\pm x$.
These represent separate excitation configurations.
Each expression is used only in its physical
propagation domain, $\sigma x\geq0$. Substitution into Eqs.~\eqref{eq:A_Ex} and
\eqref{eq:A_Ez} gives
\begin{align}
E_{x,\sigma}(x,z)
&=-\frac{iA_\sigma h'(z)}
{\omega\varepsilon_0\varepsilon(z)}
e^{i\sigma\beta x},
\label{eq:A_directional_Ex}\\
E_{z,\sigma}(x,z)
&=-\frac{\sigma\beta A_\sigma h(z)}
{\omega\varepsilon_0\varepsilon(z)}
e^{i\sigma\beta x}.
\label{eq:A_directional_Ez}
\end{align}
The directional electric field is therefore
\begin{equation}
\begin{split}
\mathbf E_{\mathrm{SPP},\sigma}(x,z)
=-\frac{A_\sigma e^{i\sigma\beta x}}
{\omega\varepsilon_0\varepsilon(z)}
\left[
ih'(z)\hat{\mathbf x}
+\sigma\beta h(z)\hat{\mathbf z}
\right].
\end{split}
\label{eq:A_directional_E}
\end{equation}
Reversal of the propagation direction changes
the sign of $E_z$, whereas $E_x$ retains the same
sign because it is determined by the transverse
derivative of the magnetic field. The fluorescent object is immersed in the upper
water at $z_0=v+a$, where $a\geq0$ is the distance
from the upper metal--water interface. At this
plane $h(z_0)=B_w e^{-\kappa_wa}$ and $h'(z_0)=-\kappa_wB_w e^{-\kappa_wa}$.
Consequently,
\begin{equation}
\mathbf{E}_{\mathrm{SPP}}(x,z_0)=C(z_0)e^{i\sigma\beta x}\left(i\kappa_w\hat{\mathbf{x}}-\sigma\beta\hat{\mathbf{z}}\right),
\label{eq:A_exterior_field}
\end{equation}
where $C(z_0)
= A_{\sigma} B_w e^{-\kappa_w a}/\omega\varepsilon_0\varepsilon_w
$.
This field includes the transmission through the finite upper metal film and the evanescent
decay in water. 
For $a=0$, the field is evaluated at $z=v^+$,
on the water side of the metal--water interface.
A finite molecular separation
can be retained by taking $a>0$. The dielectric response of the metallic layers is described using the experimentally measured optical constants of Johnson and Christy \cite{johnson_optical_1972}, from which the complex refractive index $\tilde n=n+ik$ is obtained. The metal permittivity then follows as $ \varepsilon_m = \tilde n^{\,2} = \varepsilon_m' + i\varepsilon_m''$, with $\varepsilon_m'=n^2-k^2$ and $\varepsilon_m''=2nk$. Because the metal permittivity is complex, the propagation constant is likewise complex, $\beta=\beta'+i\beta''$. The real and imaginary parts of $\beta$ are obtained by numerically solving Eq.~\eqref{eq:A_dispersion} at each frequency. 
As an illustration, Table~\ref{tab:propagation_constants1} lists representative values of the complex propagation constant $\beta=\beta'+i\beta''$ for the finite water-clad metal--dielectric--metal structure studied here, with an idealized lossless spacer of permittivity $\varepsilon_d=2.25$ ($n=1.5$).
The large values of $\beta'$ reflect strong subwavelength confinement, while the finite value of $\beta''$ determines the propagation length and ultimately limits the attainable resolution enhancement.

In the device of Fig.~3 of the main text, the gap plasmon is launched by a grating coupler in the lower Ag film; tapered input channels, as used for gap-plasmon nanofocusing~\cite{Choo2012}, offer an alternative.

\begin{table}[ht!]
\caption{Representative values of the complex propagation constant
$\beta=\beta'+i\beta''$ and the intensity propagation length
$L_{\mathrm{SPP}}=1/(2\beta'')$ for the even TM gap-plasmon mode
supported by the finite water--metal--dielectric--metal--water
structure. The outer water cladding and dielectric spacer have
permittivities $\varepsilon_w=1.77$ and $\varepsilon_d=2.25$,
respectively.}
\label{tab:propagation_constants1}
%\centering
\begin{ruledtabular}
\begin{tabular}{cccccc}
%\hline\hline
$E$ (eV) &
$\beta'$ (\textmu{}m$^{-1}$) &
$\beta''$ (\textmu{}m$^{-1}$) &
$t_d$ (nm) &
$t_m$ (nm) &
$L_{\mathrm{SPP}}$ ($\mu\mathrm{m}$) \\
%\hline
1.02 & 21.3  & 0.164 & 10 & 20  & 3.05  \\
2.01 & 100.0 & 2.07  & 3  & 20  & 0.242 \\
2.01 & 63.4  & 1.00  & 6  & 20  & 0.500 \\
2.01 & 48.2  & 0.663 & 10 & 20  & 0.754 \\
2.01 & 30.4  & 0.344 & 30 & 20  & 1.45  \\
2.01 & 44.3  & 0.569 & 10 & 30  & 0.879 \\
2.01 & 42.6  & 0.519 & 10 & 100 & 0.963 \\
2.50 & 69.0  & 1.27  & 10 & 20  & 0.394 \\
%\hline\hline
\end{tabular}
\end{ruledtabular}
\end{table}

\section{Derivation of the normalized
external-field--SPP excitation intensity}
\label{app:water_intensity}
This section derives the excitation intensity produced by
coherent interference between a directional SPP mode and an
applied optical field with matching spatial attenuation and
complex polarization.  We use the modal fields and fluorophore
position defined in Sec.~\ref{app:water_mdm}.
At the fluorophore plane, the directional SPP field is written as
\begin{equation}
\mathbf{E}_{\mathrm{SPP}}(x,z_0)
=
\mathbf{u}(z_0)e^{i\beta'x}e^{-\beta''x},
\qquad x\geq0,
\label{eq:B_spp}
\end{equation}
where $\mathbf{u}(z_0)
=
C(z_0)
\left(
i\kappa_w\hat{\mathbf{x}}
-
\beta\hat{\mathbf{z}}
\right)$ and $C(z_0)
= A B_w e^{-\kappa_w a}/\omega\varepsilon_0\varepsilon_w$.
Here, $A$, $B_w$, and $\kappa_w$ are defined in
Sec.~\ref{app:water_mdm}. We denote the local components of
$\mathbf{u}$ by $u_x=i\kappa_w C(z_0)$ and $u_z=-\beta C(z_0)$ and define $P(z_0)
=
\mathbf{u}\cdot\mathbf{u}^*
=
|u_x|^2+|u_z|^2$. The coherent applied field is specified at the same plane as
\begin{equation}
\mathbf{E}_a(x)
=
E_0g_a(x)e^{i(k_a x+\phi)}\mathbf{p}_a,
\label{eq:B_applied}
\end{equation}
Here, $E_0$ is the local applied-field amplitude at $x=0$,
$g_a(x)$ is a real, nonnegative, dimensionless amplitude
envelope normalized by $g_a(0)=1$, 
$k_a=k_0\sqrt{\varepsilon_w}\sin\alpha$ denotes the
lateral phase gradient, and $\phi$ is the controlled relative
phase. We initially retain a general envelope and subsequently
consider the matched-attenuation case,
$g_a(x)=e^{-\beta''x}$ for $x\geq0$. The unit polarization vector is
\begin{equation}
\mathbf{p}_a
=
\cos\theta\,\hat{\mathbf{x}}
+
e^{i\gamma}\sin\theta\,\hat{\mathbf{z}},
\qquad
\mathbf{p}_a\cdot\mathbf{p}_a^*=1.
\end{equation}
The angle $\theta$ determines the relative component amplitudes,
while $\gamma$ determines their relative phase and 
permits the applied field to
reproduce the phase difference between the $x$ and $z$
components of the SPP field.
Physically, $\mathbf{E}_a$ is the field of a $p$-polarized plane wave incident from the water side at angle $\alpha$ in the $x$--$z$ plane.
Because the incident and reflected waves share the in-plane wavevector $k_a$, their total field at $z_0=v+a$ 
retains the form of Eq.~\eqref{eq:B_applied}, with
\begin{equation}
\begin{aligned}
&\mathbf{p}_a\propto\big(\cos\alpha\,[1-\tilde r_p],\;\sin\alpha\,[1+\tilde r_p]\big),\\
&\tilde r_p=r_p\,e^{2ik_z a},\qquad k_z=k_0\sqrt{\varepsilon_w}\cos\alpha.
\end{aligned}
\label{eq:B_pa_phys}
\end{equation}
Here $r_p(\alpha)$ is the reflection coefficient of the multilayer for $H_y$, referred to the
upper metal--water interface $z=v$. It follows from the same boundary conditions as in Sec.~\ref{app:water_mdm},
with a real in-plane wavevector $k_a$ and incident plus reflected waves in the upper water region.
Equation~\eqref{eq:B_pa_phys} gives $\tan\theta\,e^{i\gamma}=\tan\alpha\,(1+\tilde r_p)/(1-\tilde r_p)$,
so $\theta$ and $\gamma$ are not independent parameters but are fixed by $\alpha$.
The complex polarization overlap is
\begin{equation}
S(z_0)
=
\mathbf{u}(z_0)\cdot\mathbf{p}_a^*
=
u_x\cos\theta+u_z e^{-i\gamma}\sin\theta.
\end{equation}
Its polar representation is
$S(z_0)=Q(z_0)e^{i\chi(z_0)}$, where
$Q=|S|$ and $\chi=\arg S$ denote its magnitude and phase,
respectively. The normalized reference polarization implies
$Q\leq\sqrt{P}$, with equality only when the applied field
and the local SPP field have the same complex polarization.
Combining the SPP and applied fields yields the excitation intensity
$I_{\mathrm{tot}}=|\mathbf{E}_{\mathrm{SPP}}+\mathbf{E}_a|^2$,
giving
\begin{eqnarray}
\hspace{-0.1cm}
I_{\mathrm{tot}}(x,z_0)
&=&
E_0^2g_a^2(x)+P(z_0)e^{-2\beta''x}+
2E_0Q(z_0)g_a(x)
\nonumber\\
&&
e^{-\beta''x}\cos\left[(\beta'-k_a)x-\phi+\chi\right].
\label{eq:B_total}
\end{eqnarray}
Here, the first term gives the applied-field contribution, while
the second describes the SPP intensity attenuated by propagation
loss. The third is the interference contribution, whose amplitude
depends on the polarization overlap and whose fringe position
is controlled by the relative phase $\phi$.
We now consider the  case in which the applied-field
amplitude has the same spatial attenuation as the SPP $g_a(x)=e^{-\beta''x}$ and we define
the spatial frequency difference $K=\beta'-k_a$ and the
relative phase $\Psi=Kx-\phi+\chi$.
The total intensity then takes the form
\begin{equation}
I_{\mathrm{tot}}(x,z_0)
=
e^{-2\beta''x}
\left[
E_0^2+P(z_0)+2E_0Q(z_0)\cos\Psi
\right].
\end{equation}
At positions satisfying $\Psi=(2n+1)\pi$, the interference
is destructive, giving
\begin{eqnarray}
I_{\min}(x,z_0)
&=&
e^{-2\beta''x}
\Bigl[
\left(E_0-\sqrt{P(z_0)}\right)^2
\nonumber\\
&+&
2E_0\left(\sqrt{P(z_0)}-Q(z_0)\right)
\Bigr].
\end{eqnarray}
Since $Q(z_0)\leq\sqrt{P(z_0)}$, both contributions in
brackets are nonnegative. Exact intensity zeros therefore
require equal local amplitudes,
$E_0=\sqrt{P(z_0)}$, and complete polarization overlap,
$Q(z_0)=\sqrt{P(z_0)}$.
For nonzero $u_x$ and $u_z$, the polarization condition requires
\begin{equation}
\hspace{-0.15cm}
\tan\theta=\frac{|u_z|}{|u_x|}
=\frac{|\beta|}{|\kappa_w|},
\hspace{0.3cm}
\gamma=\arg\!\left(\frac{u_z}{u_x}\right)
=\arg\left(\frac{i\beta}{\kappa_w}\right).
\label{eq:B_match}
\end{equation}
With Eq.~\eqref{eq:B_pa_phys}, this becomes a condition on the tilt angle,
\begin{equation}
\tan\alpha\,\frac{1+\tilde r_p}{1-\tilde r_p}=\frac{i\beta}{\kappa_w}.
\label{eq:B_match_alpha}
\end{equation}
A single $p$-polarized plane wave in water has a real ratio $E_z/E_x=\mp\tan\alpha$, i.e., $\gamma\in\{0,\pi\}$. 
The phase difference required by Eq.~\eqref{eq:B_match} therefore arises only from the interference between the incident and reflected waves. 
For a highly reflecting film, $\tilde r_p\approx e^{i\psi}$ with $\psi=\arg r_p+2k_za$, and the left-hand side of Eq.~\eqref{eq:B_match_alpha} becomes $i\tan\alpha\cot(\psi/2)$. 
The $z$ component of the applied field is thus phase-shifted by $\pi/2$ relative to its $x$ component, as in the SPP field, and
the magnitudes match at
\begin{equation}
\tan\alpha\approx\left|\frac{\beta}{\kappa_w}\right|\tan\frac{\psi}{2}.
\end{equation}
Since $\cot(\psi/2)>0$ here, Eq.~\eqref{eq:B_match_alpha} requires $\alpha>0$, i.e., a tilt along the SPP propagation direction.
Because $\alpha$ is a single real parameter whereas Eq.~\eqref{eq:B_match_alpha} is a complex condition, it can be
satisfied only approximately; the residual mismatch is quantified using the full-wave simulations of Fig.~3 of the main text, which show an average fringe visibility of 0.96.

In the ideal matched case, the polarization vector reproduces the local SPP field up to its magnitude and overall phase $\mathbf{u}(z_0)=\sqrt{P(z_0)}\,e^{i\chi}\mathbf{p}_a$, where $\chi=\arg u_x$. The intensity therefore reduces to
\begin{eqnarray}
I_{\mathrm{tot}}(x,z_0)
&=&
4P(z_0)e^{-2\beta''x}
\cos^2\left(\frac{\Psi}{2}\right),
\end{eqnarray}
where we used the trigonometric identity $1+\cos\Psi=2\cos^2(\Psi/2)$. We parameterize the applied-field phase as
$\phi=\chi+\zeta$, where $\zeta$ is the controlled
phase shift.
The excitation intensity becomes
\begin{equation}
I_{\mathrm{tot}}(x,z_0)
=
4P(z_0)e^{-2\beta''x}
\cos^2\left(\frac{Kx-\zeta}{2}\right).
\end{equation}
For $K\ne0$, the intensity nodes satisfy
\begin{equation}
x_n=\frac{\zeta+(2n+1)\pi}{K},
\qquad n\in\mathbb{Z},
\end{equation}
where only positions with $x_n\geq0$ are retained.
The fringe period is $\Lambda=2\pi/|K|$.
Because matching requires $k_a>0$, $K<\beta'$ and $\Lambda>2\pi/\beta'$.
For other values of $\zeta$, the fringe phase shifts while
the attenuation envelope remains fixed.

\section{Cross-sectional field maps, coupling efficiency, and the corner of the field of view}
\label{sec:fieldmaps}

\begin{figure}[h!]
\includegraphics{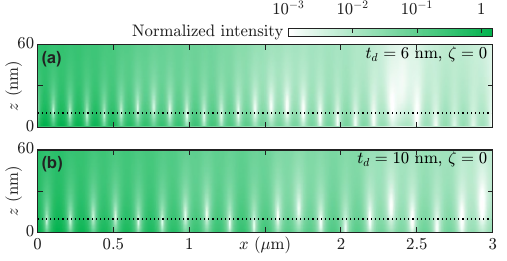}
\caption{Side view of the excitation pattern from the full-wave simulations. Normalized intensity (log scale) in the water above the top Ag film for $\zeta=0$ for (a) $t_d=6$~nm and (b) $t_d=10$~nm. Coordinate $x$ is measured from the end of the taper. Dotted lines: $z=10$~nm, the height of the line profiles in Fig.~3 of the main text; $z$ is measured from the upper metal surface.}
\label{fig:crosssec}
\end{figure}

Figure~\ref{fig:crosssec} shows the excitation pattern in the $xz$ plane from the full-wave simulations. The fringes are confined to the vicinity of the top Ag surface: the gap-plasmon field decays into the water over 16~nm ($t_d=6$~nm) and 22~nm ($t_d=10$~nm), whereas the reference beam does not, so that the fringe visibility drops from 0.99 at 10~nm to 0.6--0.8 at 30~nm above the metal surface. For $t_d=6$~nm, the residual surface plasmon on the Ag/water interface, which decays more slowly both along the surface and into the water, becomes significant beyond $x\approx2$~\textmu{}m.

For the grating couplers of Fig.~3 of the main text, the coupling efficiency into the gap plasmon at the start of the field of view, including the propagation loss under the Cr-capped overlayer, is 0.13\% for $t_d=6$~nm and 0.04\% for $t_d=10$~nm. An incident power of about 0.6~mW ($t_d=6$~nm) and 1.4~mW ($t_d=10$~nm) per micrometer of coupler length thus gives a peak excitation intensity of 1~kW/cm$^2$ at the start of the field of view. The couplers were designed for a clean excitation pattern rather than for efficiency, which could be increased by optimizing the grating.
In an experiment, the gap plasmon and the reference beam can be derived from the same laser, with the phase $\zeta$ set by a phase modulator in the reference arm. 
The exponential envelope of the reference amplitude and the field stop can be imposed by a spatial light modulator in a plane conjugate to the sample. The local fringe phases can be calibrated by fitting the phase-stepped photon counts of sparse immobilized fluorophores at the sample height, using positions estimated from the summed frames, as in SIMFLUX~\cite{Cnossen2020}.

\begin{figure}[t]
\includegraphics{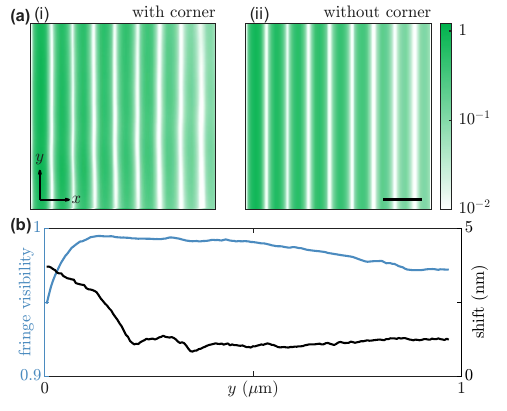}
\caption{Corner where the $x$- and $y$-oriented couplers meet, from the three-dimensional simulation for $t_d=6$~nm. (a) Normalized intensity (log scale) for $\zeta=0$ over $0.96\times0.96$~\textmu{}m$^2$ (i) with the corner and (ii) without it (two-dimensional simulation); the launching coupler lies beyond the left edge and the orthogonal coupler beyond the bottom edge of each map. Scale bar: 200~nm. (b) Fringe visibility (left axis, blue) and fringe displacement relative to the pattern without the corner (right axis, black) as functions of the distance $y$ from the end of the orthogonal coupler's taper (median over $x>50$~nm).}
\label{fig:corner}
\end{figure}

To assess the corner where the $x$- and $y$-oriented couplers meet, we solved the launch field of the $t_d=6$~nm device with a three-dimensional volume integral equation that uses the Green's tensor of the layered stack~\cite{PaulusMartin2001}, so that the only unknowns are the fields in the glass-filled slits of both gratings and in the upper metal around the corner;
far from the corner, the launching arm is continued by the two-dimensional solution, and the resulting excitation patterns are converged to within 0.3\%.
The reference beam is approximated by its field on the planar stack, so its scattering by the corner is not included.
Figure~\ref{fig:corner} compares the result with the two-dimensional pattern without the corner. Next to the orthogonal coupler, the median fringe visibility drops from 0.985 to 0.95 and the fringes are displaced by up to $\sim$4~nm; beyond 0.2~\textmu{}m, the visibility remains above 0.97 and the displacement is below 1.5~nm. The corner thus affects only a narrow band along the edges of the FOV.

\section{Cram\'er-Rao bound for localization}
\label{app:crb}

To benchmark the performance of PIFLUX and compare it with other techniques, we compute the Cram\'er-Rao bound (CRB) from the Fisher information (FI) \cite{moon2000mathematical,chao_fisher_2016}.
The emission point-spread function (PSF) is modeled using a scalar Fourier-optics description,
\begin{equation}
\label{eqn:psf}
\text{PSF}(x,y)=A\left|\mathcal{F}\left\{\text{Circ}\left(\frac{n\rho}{\text{NA}}\right)\frac{\exp\!\left[i k_\text{em}(x u+y v)\right]}{\left(1-\rho^2\right)^{1/4}}\right\}\right|^2,
\end{equation}
where $\mathcal{F}$ denotes a 2D Fourier transform, $(u,v)$ are pupil-plane coordinates, $\rho=\sqrt{u^2+v^2}$, and $k_\text{em}=2\pi n/\lambda_{\mathrm{em}}$. 
$\mathrm{Circ}(\cdot)$ is the circular pupil function set by the objective NA (with cutoff $\rho\le \mathrm{NA}/n$), and $A$ is a normalization factor chosen such that the PSF integrates to unity. We use a water-immersion objective with $\mathrm{NA}=1.2$ and $n=1.33$, $\lambda_{\mathrm{em}}=676$~nm, and an effective pixel size of 65~nm in the sample plane.

For an SM at position $(x,y)$ with total signal $s$ and total background $b$ per pixel (both summed over $N$ frames: $N=6$ for one-sided PIFLUX and SIMFLUX, $N=12$ for two-sided PIFLUX and iterative MINFLUX, $N=4$ for MINFLUX, and $N=1$ for conventional SMLM), we generated synthetic data under an independent Poisson noise model. Specifically, for frame $m$ the detected image is sampled as
\begin{equation}
I_{\mathrm{em},m}(x,y)\sim \mathrm{Poisson}\!\left(s\, I_{\mathrm{ex},m}(x,y)\,\mathrm{PSF}(x,y) + b/N\right),
\label{eqn:noiseModel}
\end{equation}
where $I_{\mathrm{ex},m}(x,y)$ is the excitation intensity profile for frame $m$, normalized such that $\sum_{m=1}^{N} I_{\mathrm{ex},m}(x,y)=1$ at the emitter position.
The FI matrix for $(x,y)$ is then
\begin{align}
\mathrm{FI}&=\nonumber\\
&\sum_{m=1}^{N}\sum_{p}\frac{1}{I_{p,m}}
\begin{bmatrix}
\left(\partial_x I_{p,m}\right)^2 & \left(\partial_x I_{p,m}\right)\left(\partial_y I_{p,m}\right)\\
\left(\partial_x I_{p,m}\right)\left(\partial_y I_{p,m}\right) & \left(\partial_y I_{p,m}\right)^2
\end{bmatrix},
\end{align}
where $I_{p,m}$ is the expected photon count in pixel $p$ of frame $m$, and $\partial_x$ and $\partial_y$ denote derivatives with respect to the emitter position. The CRB is then given by the inverse Fisher information, $\text{CRB}=\text{FI}^{-1}$.

The normalization of $I_{\mathrm{ex},m}$ fixes the number of detected photons at every position. With two-sided excitation and a fixed excitation power, the excitation summed over all frames at the center of the field of view is 1/13 ($t_d=6$~nm) and 1/11 ($t_d=10$~nm) of that at its corners for the full-wave patterns, so emitters there need correspondingly longer to emit the same number of photons.

Averaged over the $D\times D$ field of view of Fig.~5(b,c) of the main text, on a uniform $12\times12$ grid of positions at each of which the CRB is averaged over the fringe phase, the CRB of two-sided PIFLUX is 1.8 and 3.0~nm for $t_d=6$~nm and 2.2 and 3.6~nm for $t_d=10$~nm at $b=1$ and 20, respectively, compared with 3.2 and 5.9~nm for SIMFLUX. PIFLUX is more precise than SIMFLUX at every sampled position: the ratio of the two precisions never exceeds 0.66 ($b=1$) and 0.58 ($b=20$) for $t_d=6$~nm, and 0.79 and 0.74 for $t_d=10$~nm.

\section{Maximum-likelihood estimation}
\label{app:mle}

We localized simulated emitters with maximum-likelihood estimators under the Poisson noise model of Eq.~\eqref{eqn:noiseModel}, with the PSF of Eq.~\eqref{eqn:psf}, 500 signal photons per localization, and a total background of $b=1$ or 20 photons per pixel distributed over all frames of the localization, as in the CRB analysis (Appendix~\ref{app:crb}).

For PIFLUX, each cycle consists of 12 frames: three phases, $\zeta=0$, $2\pi/3$, and $4\pi/3$, for each axis and each of the two launching couplers. An emitter emits at a rate proportional to its local excitation intensity, with the excitation power set such that an emitter at the origin emits its 500 photons within one cycle; emitters elsewhere spread their photons over several cycles. The estimator proceeds in three stages. First, all frames are summed and cross-correlated with the PSF to locate the emitter to within a pixel and to define a $7\times7$-pixel region of interest (ROI). This ROI encloses most of the PSF, whose full width at half maximum is $\sim$290~nm, while limiting the background. Smaller ROIs clip the PSF, and larger ones did not improve the precision. Second, for each axis, we maximize the Poisson likelihood of the photon counts in the ROI of the frames patterned along that axis over 500 positions spanning $\pm$half a fringe period (capped at one pixel) around the coarse position. The signal amplitude is set at each candidate position by the background-subtracted total count, so the photon number is not assumed known. Third, the search is repeated over 500 positions spanning a ten times narrower range around the coarse maximum. The two per-axis estimates form the 2D localization.

SIMFLUX uses the per-pixel maximum-likelihood estimator of Ref.~\cite{Cnossen2020} on all six frames, with the pattern (pitch 257~nm, full modulation) and the background known. MINFLUX uses three iterations with $L=100$, 50, and 25~nm and one third of the photons each, with four exposures per iteration and a residual intensity of 0.5\% at the donut zero. The first iteration is centered on the true emitter position, a best case for MINFLUX, and each later one on the previous estimate; the final estimate maximizes the summed likelihood of the three iterations.

\begin{figure}[ht!]
\centering
\includegraphics[width=\linewidth]{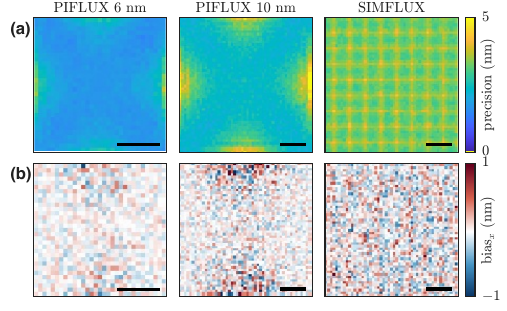}
\caption{Precision and accuracy of the estimators across the FOV for $b=1$. (a) Localization precision $\sigma_r$ and (b) mean localization error along $x$ for PIFLUX with $t_d=6$~nm (FOV $3\times3$~\textmu{}m$^2$) and $t_d=10$~nm ($5\times5$~\textmu{}m$^2$), and for SIMFLUX ($5\times5$~\textmu{}m$^2$), from 200 localizations with 500 photons at each point of a 100-nm grid. The $y$ error follows from the $x-y$ symmetry. Scale bars: 1~\textmu{}m.}
\label{fig:mle_b1}
\end{figure}

\begin{figure}[ht!]
\centering
\includegraphics[width=\linewidth]{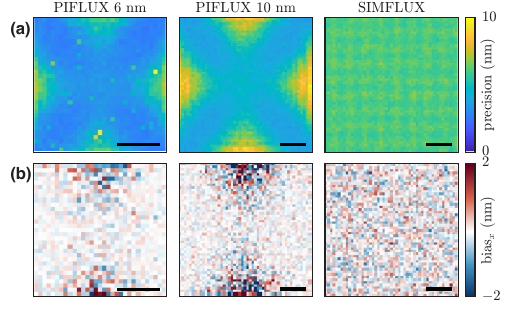}
\caption{Precision and accuracy of the estimators across the FOV for $b=20$. (a) Localization precision $\sigma_r$ and (b) mean localization error along $x$ for PIFLUX with $t_d=6$~nm (FOV $3\times3$~\textmu{}m$^2$) and $t_d=10$~nm ($5\times5$~\textmu{}m$^2$), and for SIMFLUX ($5\times5$~\textmu{}m$^2$), from 200 localizations with 500 photons at each point of a 100-nm grid. The $y$ error follows from the $x-y$ symmetry. Scale bars: 1~\textmu{}m.}
\label{fig:mle_b20}
\end{figure}

Figures~\ref{fig:mle_b1} and \ref{fig:mle_b20} show the precision and accuracy of the estimators across the FOV, from 200 localizations at each point of a 100-nm grid. Each emitter switches on at the start of a cycle. At $b=1$, the median precision $\sigma_r$ is 1.8~nm for PIFLUX with $t_d=6$~nm, 2.4~nm with $t_d=10$~nm, and 3.3~nm for SIMFLUX, and the mean error along $x$ stays below 0.5~nm at 95\% of the positions for all three. The precision of PIFLUX degrades near the midpoint of each coupler, where the frames patterned along the coupler are excited only weakly and receive few photons. At $b=20$, this produces the X-shaped patterns in Fig.~\ref{fig:mle_b20}, where the fringe order is occasionally misassigned; the median $\sigma_r$ is then 3.3~nm, 4.5~nm, and 6.1~nm, respectively. The SIMFLUX maps show a beat pattern because its precision depends on the fringe phase, which differs between neighboring grid points because the 100-nm grid spacing is incommensurate with the 257-nm pitch.
Averaged over the field of view, the precision is 1.9~nm, 2.6~nm, and 3.3~nm at $b=1$, and 3.7, 4.9, and 6.1~nm at $b=20$ for PIFLUX with $t_d=6$~nm, $t_d=10$~nm, and SIMFLUX, respectively.

For the NPCs of Fig.~6 of the main text, the NPCs were placed at random positions at least 250~nm apart and entirely inside the FOV with random orientations (9 per $3\times3$~\textmu{}m$^2$ FOV for $t_d=6$~nm and 25 per $5\times5$~\textmu{}m$^2$ FOV for $t_d=10$~nm); each emitter switched on at a random time within the cycle.
The PIFLUX estimator reaches the CRB evaluated over the same ROI with unknown photon number within the statistical uncertainty of the simulations; its localization error is 11--19\% ($t_d=6$~nm) and 16--28\% ($t_d=10$~nm) larger than the full-image CRB of Fig.~5 of the main text.

\begin{figure}[t!]
\centering
\includegraphics[width=\linewidth]{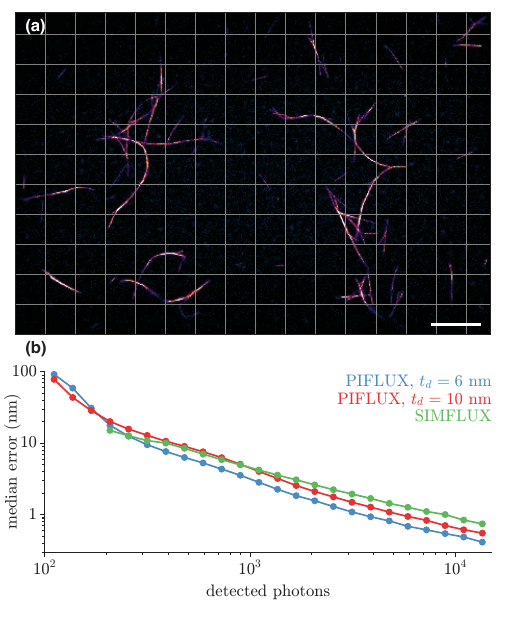}
\caption{Fibril test with experimental blinking statistics. (a) PIFLUX reconstruction ($t_d=6$~nm) of the $\alpha$-synuclein fibril dataset of Ref.~\cite{bruggeman_polcam_2024}, with the field tiled into $3\times3$~\textmu{}m$^2$ FOVs (grey lines). Scale bar: 5~\textmu{}m. The apparent fibril width of $\sim$40~nm is set by the localization uncertainty of the experimental dataset used as the ground truth, not by PIFLUX. (b) Median radial localization error versus the number of detected photons used by each localization; bins with fewer than 100 blinks are omitted.}
\label{fig:fibril}
\end{figure}

To test PIFLUX under experimental blinking statistics, we took the emitter positions and photon counts from a PAINT dataset of $\alpha$-synuclein fibrils~\cite{bruggeman_polcam_2024} and used its localizations as the ground truth.
Localizations in consecutive frames within 100~nm of each other were linked into 504,457 blinks, each with its photon-weighted position and summed photon count. Each blink was simulated in isolation. The $47.6\times32.2$~\textmu{}m$^2$ field was tiled into 176 FOVs of $3\times3$~\textmu{}m$^2$ ($t_d=6$~nm) or 70 FOVs of $5\times5$~\textmu{}m$^2$ ($t_d=10$~nm).  
Each blink lasts for a finite on-time, given by its photon count divided by an emission rate drawn from the experimental data. The blink starts at a random time within the pattern cycle.
The camera runs at 290 frames per second, so that a 12-frame PIFLUX cycle takes 41.4~ms, and the measured background of 29 photons per pixel per localization is distributed over the frames used.

Figure~\ref{fig:fibril}a shows the resulting PIFLUX reconstruction; Fig.~\ref{fig:fibril}b compares the median localization error at the same number of detected photons. With $t_d=6$~nm, PIFLUX is more precise than SIMFLUX above $\sim$300 photons, by a factor that grows from $\sim$1.3 at 400 photons to $\sim$1.8 above 3000 photons; with $t_d=10$~nm, it is more precise above $\sim$1000 photons. Because the blinks are short, some end before completing a 12-frame cycle and are localized less reliably; cycling the patterns faster than the on-time of the emitters would therefore further improve PIFLUX.

\bibliography{references}